# Effect of hydrostatic pressure on elastic properties of ZDTP tribofilms

Karim DEMMOU, Sandrine BEC* and Jean-Luc LOUBET

Laboratoire de Tribologie et Dynamique des Systèmes - UMR CNRS 5513
Ecole Centrale de Lyon - 36 avenue Guy de Collongue - 69134 ECULLY Cedex – France
*corresponding author: sandrine.bec@ec-lyon.fr

**Abstract**

Previous studies have shown that the elastic properties of Zinc Dialkyl-dithiophosphate (ZDTP) tribofilms measured by nanoindentation increase versus applied pressure (Anvil effect) [1, 2]. The aim of this paper is to demonstrate that, up to 8 GPa, this increase is a reversible phenomenon.

A ZDTP tribofilm has been produced on "AISI 52100" steel substrate using a Cameron-Plint tribometer. After its formation, a hydrostatic pressure of about 8 GPa was applied during one minute on the tribofilm using a large radius steel ball ("Brinell-like" test). Nanoindentation tests were performed with a Berkovich tip on pads in order to measure and compare the mechanical properties of the tribofilm inside and outside the macroscopic plastically deformed area. Careful AFM observations have been carried out on each indent in order to take into account actual contact area.

No difference in elastic properties was observed between the two areas: tribofilm modulus and pressure sensitivity are the same inside and outside the residual hemispherical print. This demonstrates that Anvil effect is a reversible phenomenon in the studied pressure range.

## Introduction

Zinc Dialkyl-dithiophosphates (ZDTP) have largely proven their efficiency as antiwear additives in car engine lubrication. It has been demonstrated that they act by forming protective films onto the rubbing surfaces under boundary lubrication conditions [3-5]. Nevertheless, because of more and more severe regulations concerning phosphorus and sulfur emissions, their use has to be drastically reduced and less pollutant new antiwear additives have to be developed. A lot of researches have been carried out on ZDTP tribofilms in order to better understand how they work. They are described in recent review papers [6, 7]. Among them, many chemical investigations have been conducted concerning film chemistry and mechanisms of film formation [8, 9] but there are few mechanical characterizations of the formed tribofilms although they can contribute to a better understanding of their mechanism of action.

The efficiency of ZDTP protective antiwear films can be related to their mechanical behavior inside the contact. Among the required properties for such films is the need to have good wear resistance, which is related to a sufficient high shear strength coupled with an adequate rate of film formation to ensure a continuous covering of the rubbing surfaces. They also need to have sufficiently low shear strength to ensure that the shearing occurs inside the film and not inside the metallic parts. Because of their patchy structure, it is extremely difficult to measure the mechanical properties of such tribofilms, inhomogeneous in thickness and structure.

Nanoindentation is an appropriate technique to measure near-surface mechanical properties of materials or mechanical properties of thin films. It has been already used to measure the hardness and Young's modulus of ZDTP tribofilms [1, 7]. The mechanical properties of a ZDTP tribofilm have also been recently measured by nanoindentation at various temperatures, from ambient temperature up to 80°C [2], and at several strain rates ($\dot{\varepsilon}$ =0.003 s$^{-1}$ to 0.1 s$^{-1}$) [10]. It has been shown that, whatever the temperature and strain-rate, both the Young's modulus of the tribofilm and its hardness increase with the applied load. In a previous work it was demonstrated that there was no gradient of properties inside the tribofilm which could explain the observed increases [1]. The tribofilm is confined between the diamond tip and the steel hard substrate and the increase of the Young's modulus of the film is due to the increase of applied pressure.

The aim of this study is to determine if this phenomenon, already known for polymeric materials as Anvil effect, is reversible or not in the case of ZDTP tribofilms.

## Methods

### Sample preparation

The tribofilm was prepared on a Cameron-Plint tribometer using a cylinder (6 mm long and 6 mm in diameter) sliding versus a plane, both in AISI 52100 steel. The cylinder slides at a frequency of 7 Hz on a distance of 7 mm. The lubricant was a synthetic poly alpha olefin (PAO) base oil with 3% in weight ZDTP (zinc bis (O,O-di-hexyl dithiophosphate)). The applied normal load was first set to 50 N during 5 minutes, then progressively increased to 350 N during a period of 5 minutes and then maintained constant (total test duration: 1 hour). The test temperature was kept constant at 80°C. After the tribological test, the sample was washed during 15 minutes in ultrasonic bath containing acetone.

### Hydrostatic pressure application

First, the macroscopic hardness of the AISI 52100 steel substrate was measured thanks to several Vickers hardness test. The average Vickers hardness of the substrate is about 7.2±0.3 GPa. The corresponding pressure obtained by dividing the indentation load by the projected residual area of indentation, is about 7.7±0.3 GPa. This value is the highest mean pressure which can be reached inside the rubbing contact with AISI 52100 steel.

Second, a ball-bearing steel ball, 17.5 mm in diameter, was used in order to perform a "Brinell-like" hardness test on the tribofilm. The ball was positioned on top of the tribofilm and a 15 kN force was applied on the ball (Figure 1).

Thus, as shown in Figure 2, the substrate was plastically deformed and, because of the print size and its flatness, the mean pressure applied on the substrate can be considered to be a hydrostatic pressure. So, the tribofilm underwent a hydrostatic pressure of about 7.7 GPa. Furthermore, the optical microscopy observation of the pads inside and outside the hemispherical print shows no visible difference between the two areas.

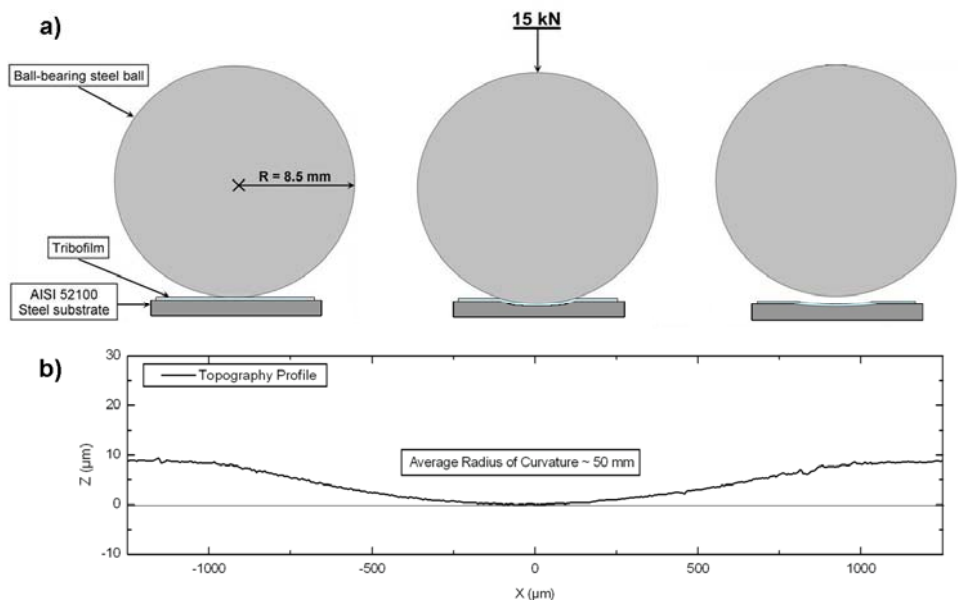

*Figure 1  a) Schematic description of the "Brinell-like" test (the thickness of ZDTP tribofilm is exaggerated).
        b) Topography profile of the residual print along a diameter. Because of its large radius of curvature, the print can be considered to be flat.*

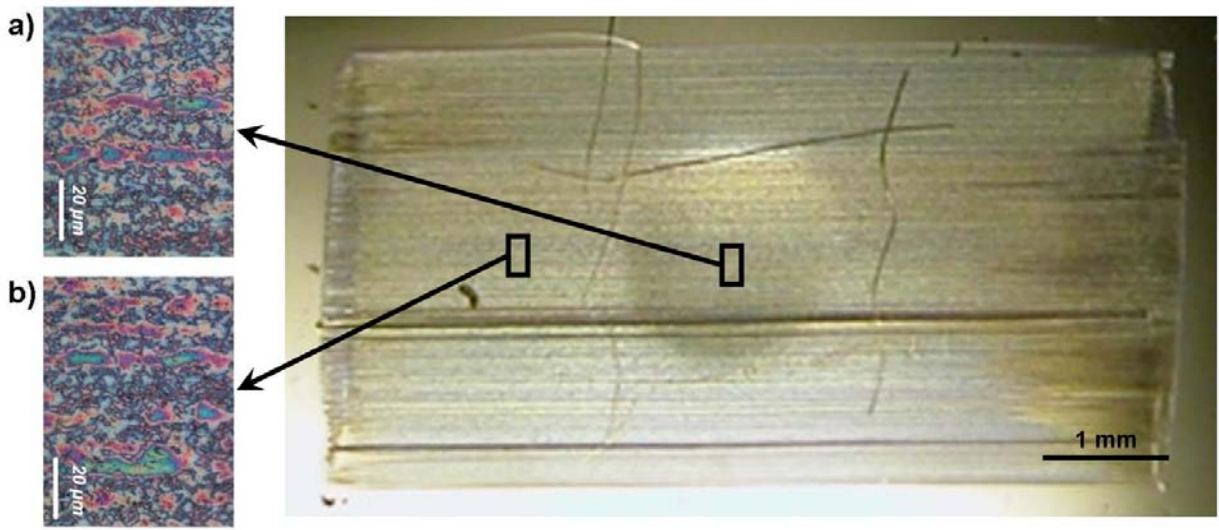

*Figure 2: Residual print in the tribofilm after the "Brinell-like" test. The residual print has a diameter of 1.8 mm. Optical observations of pads: (a) inside the "Brinell-like" residual print, (b) outside the "Brinell-like" residual print. The two areas are similar.*

*Nanoindentation tests*

Nanoindentation tests were performed with a MTS Nanoindenter SA2® equipped with the DCM (Dynamic Contact Module) from MTS systems and using the continuous stiffness measurement (CSM) method [11]. This method consists in superimposing a small displacement oscillation (small enough to generate only an elastic strain) at a given frequency during the indentation test. So, the sample is subjected to small loading-unloading cycles. The simultaneous measurement of the normal force, $F$, the contact stiffness, $S$, and the displacement into surface, $h$, allows to scan mechanical properties, hardness, $H$, and reduced Young's modulus, $E^*$, continuously along the whole indentation period. The small superimposed displacement had amplitude of 1 nm at a frequency of 75 Hz. For each test, the maximum applied load was 10 mN. With the DCM, load and displacement resolution during indentation tests are respectively < 1 nN and < 1 pm.

Using the optical microscope integrated in the device, the indentation location is precisely chosen and the position of each residual indent is verified. Inside the "Brinell-like" hemispherical print, the indentations were carried out on pads included in a circle of radius 250 µm centered in the center of the residual mark in order to guarantee a satisfactory flatness. Nanoindentation tests were conducted at a constant strain rate, $\dot{\varepsilon} = 0.003$ s$^{-1}$ [12].

*AFM observations*

A VEECO® CP-II Atomic Force Microscope was used to perform observations of the indents and particularly of the pile-up at their periphery.

Because of the very low load (10 mN), the residual indents were about 2 or 3 µm in size. AFM permitted topography measurements with high three dimensional resolution. Thereby, pads thickness could be estimated and pile-up around the indents could be quantified in order to obtain reliable hardness values considering actual contact area.

AFM was used in contact mode with sharp silicon probes. For each indent, large observations of the whole pad, observation of the residual indent itself and detailed image of the pile-up around the indent were carried out. In order to perform actual contact area measurements, the "error mode" images were chosen. The "error signal" is similar to the derivation of the topography signal and offers a better contrast than the height image regarding slope changes and it is more sensitive to determine the actual contact area.

## Results and discussion

Figure 3 shows AFM topography and "error mode" images obtained for two nanoindentation tests performed inside and outside the "Brinell-like" residual print. The "error mode" images were used to measure accurately the actual contact area taking into account the pilled-up matter around the indent [2]. Larger topography AFM images were also used to estimate the thickness of the pads on which nanoindentation tests were performed. This confirms that there is no difference in pad's shape and height between pads inside outside the "Brinell-like" residual print. The pad's thicknesses were measured between 150 and 250 nm both inside and outside the residual hemispherical print. A simple model was then used to take into account the effect of the substrate on the measured elastic properties in order to extract the elastic modulus of the tribofilm itself [13].

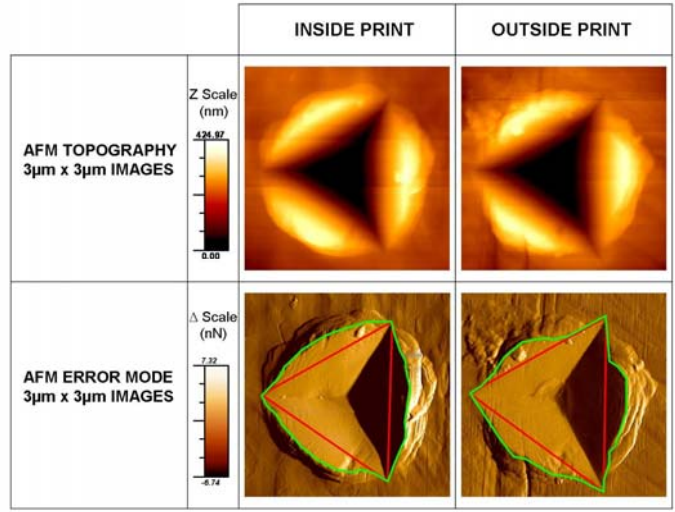

*Figure 3: AFM observations corresponding to indentation experiments performed inside and outside the "Brinell-like" hemispherical residual print. AFM images show no difference in indent morphology for these two tests. AFM error mode images allow measurement of the actual contact area (green polyhedron), which is different from the theoretical contact area (red triangle).*

It was observed that the Young's modulus of the tribofilm increased with the applied pressure. This phenomenon, called "Anvil effect", has been observed previously during nanoindentation of thin polymer films on hard substrate [14] and was also systematically observed for ZDTP tribofilms [1, 2]. For the tribofilm, "Anvil effect" takes place from a threshold pressure value, which is the initial measured hardness $H_0$ (Figure 4). From this threshold hardness value, the film's Young modulus increases from its initial value, $E^*_{f0}$, proportionally to hardness.

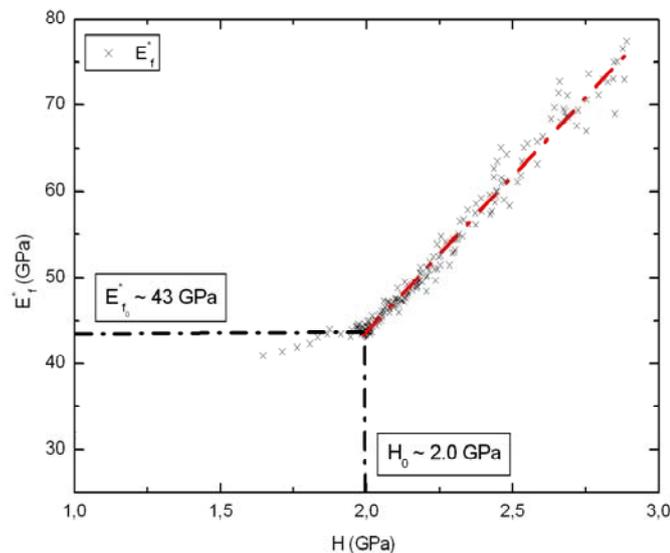

*Figure 4: Representative curve of tribofilm's reduced modulus versus measured hardness. The initial film's modulus $E^*_{f0}$ is about 43 GPa and increases with increasing hardness from a threshold hardness value $H_0$ of approximately 2.0 GPa.*

Figure 5a shows two examples of nanoindentation curves obtained inside and outside the "Brinell-like" residual print. Figure 5b highlights the mean evolution of the tribofilm's reduced modulus, $E^*_f$, versus indentation plastic depth, $h_p$, for nanoindentation tests performed inside and outside the hemispherical residual print. Each curve is the average of 5 nanoindentation tests performed on the top of pads of comparable thickness (~250 nm). The two curves have similar shape. The difference between the initial moduli $E^*_{f0}$ for the two curves is within the measurement uncertainty (about 10%). This leads to a mean $E^*_{f0}$ value of 41±5 GPa, in good agreement with previous studies [1, 2].

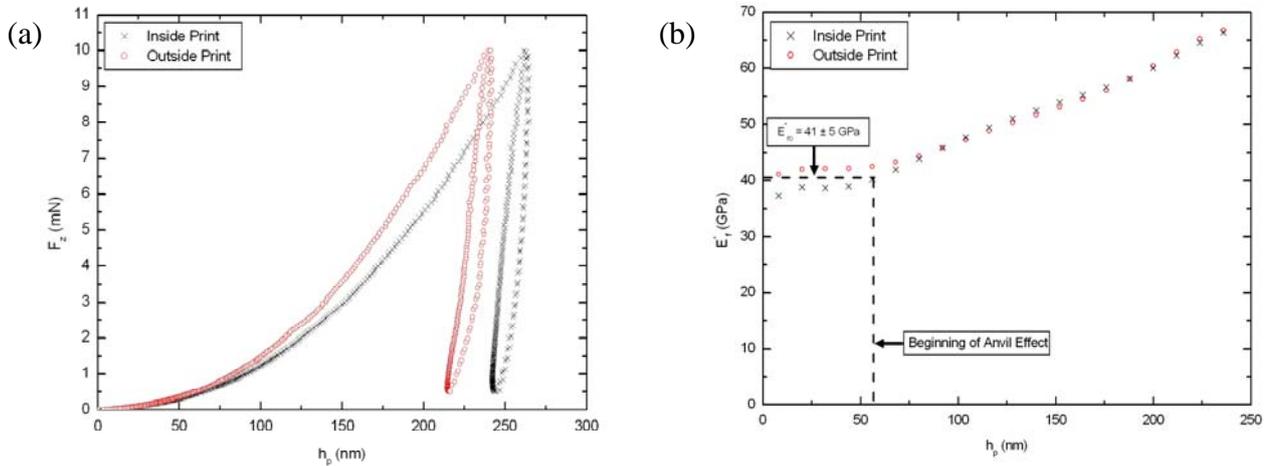

*Figure 5: (a) Examples of force-depth curves for the indents shown in figure 3. (b) Reduced modulus versus indentation depth inside (red circle) and outside (black cross) the "Brinell-like" residual print. Each curve is an average curve obtained from 5 indentation tests performed on different pads.*

So, this demonstrates that, in the studied pressure range, i.e. between 2 and 7.7 GPa, the effect of the applied pressure on the reduced modulus of the tribofilm is reversible. Since elastic properties of a material are linked to atomic bonding, it can be concluded that, applying a pressure lower than approximately 8 GPa, does not lead to irreversibly change bondings inside the tribofilm. This is in good agreement with the results obtained by Gauvin et al. [15]. In their study, model materials (zinc orthophosphate and amorphous and crystalline zinc pyrophosphate) were pressurized in a diamond anvil cell and Raman spectroscopy was used to observe the effects of high hydrostatic pressure on zinc-phosphate chains length. The results show no evidence for polymerization of phosphate compounds when pressurized up to 20.7 GPa: when subjected to high pressure, short-chain zinc pyro- and ortho-phosphate do not polymerize whether they are in the glassy or crystalline form.

Moreover, if the Anvil effect intensity is defined as the slope of the increase of $E^*_f$ versus increasing pressure, it is the same whatever the hydrostatic pressure applied on the pads before nanoindentation tests. The measured value of Anvil effect intensity is about 25 ± 10%. This shows that the sensitivity of tribofilm modulus to hydrostatic pressure, in the studied pressure range, is not influenced by the hydrostatic pressure history of the tribofilm.

**Conclusions**

Thanks to a "Brinell-like" test, a high hydrostatic pressure, about 8 GPa, was applied on a ZDTP tribofilm. Comparison between nanoindentation tests performed inside and outside the residual print of the "Brinell-like" test pointed out no differences concerning elastic properties and Anvil effect. This demonstrates that the increase of Young's modulus observed on ZDTP tribofilms during the nanoindentation test is a reversible phenomenon in the studied pressure range (2-8 GPa).

**Acknowledgements**

The authors would like to thank C. Minfray for collaboration and Asahi Denka Co., Ltd. for providing pure zinc dialkyl dithiophosphate used in these experiments.